\documentclass[12pt]{article}
\usepackage{epsfig}

\def\beq{\begin{equation}}
\def\eeq#1{\label{#1}\end{equation}}
\def\eeqn{\end{equation}}

\def\beqa{\begin{eqnarray}}
\def\eeqa#1{\label{#1}\end{eqnarray}}
\def\eeqan{\end{eqnarray}}

\let\bar=\overbar

\def\D{{\cal D}}

\def\O{{\cal O}}

\def\Dslash{\not{\hbox{\kern-4pt $D$}}}
\def\dslash{\not{\hbox{\kern-2pt $\del$}}}

\def\msb{{\bar{\ssstyle M \kern -1pt S}}}

\def\Title#1{\begin{center} {\Large {\bf #1} } \end{center}}

\begin{document}
\newcommand{\mx} {\ensuremath{m_{X}}\xspace}
\newcommand{\mxqsq} {\ensuremath{(m_{X}, q^2)}\xspace}
\newcommand {\pplus}  {\ensuremath{P_{+}}\xspace}
\newcommand{\elsmax}{\ensuremath{(E_{\ell},s_{\mathrm{h}}^{\mathrm{max}})}}
\newcommand{\smax}{\ensuremath{s_{\mathrm{h}}^{\mathrm{max}}}}
\newcommand{\el} {\ensuremath{E_{\ell}}\xspace}
\newcommand {\mb}{\ensuremath{m_b}}
\newcommand {\mc}{\ensuremath{m_c}\xspace}
\newcommand{\btou}{\ensuremath{\bar{B} \to X_u\, l\,  \bar{\nu}_l}}
\newcommand{\btoc}{\ensuremath{\bar{B}\to X_c\, l\,  \bar{\nu}_l}}
\newcommand{\bsg}{\ensuremath{\bar{B}\to X_s\, \gamma}}

\newcommand{\Vcs} {\ensuremath{|V_{cs}| }}
\newcommand{\Vcd} {\ensuremath{|V_{cd}| }}
\newcommand{\Vcb} {\ensuremath{|V_{cb}| }}
\newcommand{\Vub} {\ensuremath{|V_{ub}| }}
\def\ra{\rightarrow}
\def\D0{D\O }

\Title{Semileptonic $B$ and $D$ decays-latest developments }

\bigskip\bigskip

\begin{raggedright}  

{\it Giulia Ricciardi \index{Ricciardi, G.}\\
Dipartimento di Scienze Fisiche,  Universit\`a di Napoli Federico II \\ Compl. Univ. di Monte Sant'Angelo, Via Cintia, I-80126 Napoli,
ITALY \\
and \\
INFN, Sezione di Napoli \\
 Compl. Univ. di Monte Sant'Angelo, Via Cintia, I-80126 Napoli,
ITALY}
\bigskip\bigskip
\end{raggedright}

\section{Introduction}

Semileptonic  $B$ and $D$  decays are on a well deserved podium to extract CKM matrix elements, to validate theoretical tools and, more recently, even to look for hints of new physics.
%Semileptonic $B$ and $D$ decays
They  provide information about the CKM matrix elements $|V_{cb}|$, \Vub, $|V_{cd}|$  and $|V_{cs}|$ through exclusive and inclusive 
processes driven by  $b \ra c(u)$ and   $c \ra s(d)$ decays, respectively.
All semileptonic $B$ and $D$ decays  share the theoretical advantages of being  tree level dominated and of  a better control of non-perturbative parameters, due to the possibility to
factorize  leptonic and hadronic currents, with respect to hadronic decays.
Inclusive and exclusive decays differ
on both theoretical and experimental grounds,  and their comparison is a useful probe into strong interaction dynamics.

Here we review the most recent updates.
Part of them is  already covered in Refs. \cite{Ricciardi:2012pf, Ricciardi:2012xu},
to  which we refer for more details. 
%To present an exhaustive list of references is outside the scope of this review, the focus being on latest developments.

%; in the following we will mostly focus
% on  topics that have not been already  discussed there.

\section{$ b \rightarrow c \, l \, \nu$ decays}

The $ b \rightarrow c $ decay rates, which are proportional  to  the CKM matrix element 
$|V_{cb}|$ squared, are a useful handle to extract  its value.
Recent data 
on $ r B \rightarrow D \, l \,  \nu$, with $l=e, \mu$,
come 
 from Babar \cite{Aubert:2008yv, Aubert:2009ac}, on
$   B \rightarrow D^{\ast} \, l \,  \nu$
from Babar \cite{Aubert:2008yv} and Belle \cite{Dungel:2010uk}.
 The 
 differential rate for  the  latter is more easily  measured,   since
the rate is more than twice and there is no background from mis-reconstructed  $B \rightarrow D^\ast \,  l \,  \nu $.
The extraction of $|V_{cb}|$ requires the knowledge of the form factors that parameterize the matrix elements.
They can be calculated by nonperturbative methods, and this is the main source of theoretical uncertainty.
In the case of lepton masses zero, we have only one form factor, which considerably simplify the theoretical approach.

In Table \ref{phidectab1} we compare  exclusive
values of $|V_{cb}|$ extracted by analyzing  $ B \rightarrow D^{(\ast)} \, l \, \nu $ decays. On the left, we list the Collaborations providing the experimental data or averages, together with the
theoretical approaches used to determinate the form factors. The results are all in agreement within the errors. The non lattice determinations give values lower thanabout  2-5 \% with respect to unquenched lattice calculations.
A debated  method to calculate form factors, alternative to lattice and QCD sum rules and not listed in  Table \ref{phidectab1},  is  the relativistic quark model based on the quasipotential approach (see, e.g., \cite{Ebert:2006nz}).

%%%%%%%%%%%%%%
\begin{table}[t]
\centering
\vskip 0.1 in
\begin{tabular}{|l|c|} \hline \hline
{\it \bf  Exclusive Decays}  &    $|V_{cb}|\,  (10^{-3})$   \\
\hline \hline
Data/Th \hspace{1cm} $ (B \rightarrow D \, l \, \nu )$   & \\
\hline
 HFAG \cite{Amhis:2012bh}/Fermilab \& MILC \cite{Okamoto}  & $  39.70 \pm 1.42_{\mathrm{exp}} \pm 0.89_{\mathrm{th}} $\\
Babar \cite{Aubert:2009ac}/lattice SSM (quenched)  \cite{de Divitiis:2007ui} &  $  41.6 \pm 1.8_{\mathrm{stat}} \pm 1.4_{\mathrm{syst}} \pm 0.7_{\mathrm{FF}} $ \\
PDG \cite{Beringer:1900zz}/non lattice BPS \cite{uraltsev} & $  40.7 \pm 1.5_{\mathrm{exp}} \pm 0.8_{\mathrm{th}}  $\\
\hline
 \hspace{2.7cm}  $ (B \rightarrow D^\ast \, l \, \nu )$    &   \\
\hline 
 HFAG  \cite{Amhis:2012bh}/Fermilab \& MILC \cite{Bailey:2010gb} & $   39.54 \pm 0.50_{\mathrm{exp}} \pm 0.74_{\mathrm{th}} $\\
HFAG  \cite{Amhis:2012bh}/Sum Rules \cite{gambino}  & $  41.6\pm 0.6_{\mathrm{exp}}\pm 1.9_{\mathrm{th}} $\\
\hline \hline
{\it \bf Inclusive Decays}  &    \\
\hline
\hline
HFAG  \cite{Amhis:2012bh} (1S scheme)  & $  41.96 \pm 0.45 $\\
\hline
HFAG  \cite{Amhis:2012bh} (kinetic  scheme)  & $ 41.88 \pm 0.73 $\\
\hline
\hline
{\it \bf Global CKM fits}  &    \\
\hline
\hline
CKMfitter  \cite{Derkach}   & $ 40.69 \pm 0.99 $\\
\hline
UTfit \cite{tar}  & $ 42.3 \pm 0.9 $\\
\hline
\end{tabular}
\caption{ Comparison of exclusive, inclusive and indirect  determinations of $|V_{cb}|$. In the exclusive section  experimental and theoretical errors are listed, respectively, except in the
 second line, where the errors are respectively statistical, systematic and due to the theoretical uncertainty in the form factor.}
\label{phidectab1}
\end{table}
%%%\end{tabular}
%%%\end{ruledtabular}
%%%\end{table}
%%%%%%%%%%%%%%%%%%%%%%%%%%%%%%%%%%%%%%

In inclusive $ B \rightarrow X_q  l  \nu$ decays, we sum over all possible final states $X_q$, no matter if single-particle  or  multi-particle states.
Since  inclusive decays  do not depend on the details of final state, quark-hadron duality is generally assumed.
In most of the phase space,  long and short distance dynamics are factorized by means of  the heavy 
quark expansion.
However, the  phase space region includes a  region of singularity, also called endpoint or threshold region, plagued by the presence of 
 of large double (Sudakov-like)  perturbative  logarithms at all orders in the strong coupling (see e.g.  \cite{ res3, res5, res6, res7}).
For  $b \rightarrow c$ semileptonic decays, the effect of the small region of singularity is not very important; in addition,  corrections are not expected  as singular as in the $ b \rightarrow u$ case, being  cut-off by the charm mass (see e.g.  \cite{resbc1, resbc3}).

In order to determine $|V_{cb}|$, a  global fit may be performed to the width and  all available measurements of moments in $ B \rightarrow  X_c \,  l \, \nu$. A global fit has been recently accomplished in both the kinetic and the 1S scheme \cite{Amhis:2012bh}. Each scheme has its own non-perturbative parameters that have been estimated  together with the charm and bottom masses.
In Table \ref{phidectab1} we report the fit results in both schemes.
The inclusive averages are higher than 
  the values extracted from exclusive decays, but in substantial agreement within the errors.
The most precise measurements are from inclusive, that are below 2\%.
Still, the determination of $|V_{cb}|$ from $\bar B \rightarrow D^\star l \bar \nu$ 
has reached the relative precision of about 2\%. 
In Table \ref{phidectab1}  we also compare with  the global fit of the CKM matrix elements within the Standard Model, as calculated by the CKMfitter and UTfit groups.
In comparing with global fits, though, one should remind that global fits are also constrained from loop  processes,   potentially 
affected by new physics, while exclusive and inclusive  studies are based on standard model tree level  
decays. Such considerations hold of course also for thel $|V_{ub}|$ extraction, discussed in 
Sect \ref{Vub}.

Until 2007,  only decays where the final lepton was an electron or a muon had  been observed.
The  first observation  of the  decay $\bar B \ra D^\ast \tau^- \bar \nu_\tau$, 
by the Belle Collaboration \cite{Matyja:2007kt}, was followed by  improved measurements and  evidence for  $\bar B \ra D \tau^- \bar \nu_\tau$,  by both Babar and Belle Collaborations
\cite{Aubert:2007dsa, Bozek:2010xy}.
%, together with
 %for $\bar B \ra D^\ast \tau^- \bar \nu_\tau$.
The measured values for \beq  {\cal R} \left(D^{(\ast)}\right)=\frac{ {\cal B} (\bar B \ra D^{(\ast)} \tau^- \bar \nu_\tau)}{ {\cal B} (\bar B \ra D^{(\ast)} l^- \bar \nu_l)} \eeq,  have been
consistently exceeding the standard model (SM) expectations. This year, 
 Babar  has updated its older  measurement \cite{Aubert:2007dsa}
by using  the full Babar data sample and 
 increasing the signal efficiency by more than a factor
of 3 \cite{Lees:2012xj}.
The resulting $ {\cal R} (D) = 0.440 \pm 0.058 \pm 0.042 $
and $ {\cal R} (D^\ast)= 0.332  \pm 0.024 \pm  0.018$   have  been  compared with the SM predictions,  updating the calculations in Refs. \cite{Fajfer:2012vx, Kamenik:2008tj}, and finding  $  {\cal R} (D)_{\mathrm SM} =   0.297 \pm
0.017$ and   $ {\cal R} (D^\ast)_{\mathrm SM} = 0.252\pm 0.003 $, averaged
over electrons and muons. The results
  exceed the standard model expectations  by $ 2.0\, \sigma$ for $ {\cal R} (D) $  and by
$ 2.7 \,  \sigma $ for  $ {\cal R} (D^\ast)$;  taken together, they  disagree at the $3.4 \, \sigma$  level  \cite{Lees:2012xj}.
In the Babar experimental analysis   \cite{Lees:2012xj},  it is   excluded that the  excess can be explained by a charged Higgs boson in the type II two-Higgs-doublet model, for any value of $\sin \beta$.

The experimental results have prompted several  theoretical studies,  where  new physics has been advocated  to take into account this disagreement. Enhancements over the standard model values have been found in  the framework of R-parity violating
 MSSM \cite{Deshpande:2012rr},  in the two Higgs doublet model of type III \cite{Crivellin:2012ye}
and in a model with four-fermi
operators having  vector/axial vector and scalar/pseudoscalar couplings \cite{Datta:2012qk}.
Other interpretations require the presence of non minimal flavour violation  right-right vector or right-left
scalar currents; also  leptoquark models or  models
with  composite  quarks and leptons  (with  nontrivial
flavor structure) have been studied \cite{Fajfer:2012jt}.
Besides checking constraints on new physics,   the theoretical inputs  leading to  the determination of   $ {\cal R} (D) $ within the standard model have also been requestioned \cite{Becirevic:2012jf, Bailey:2012jg}.

\section{$ b \rightarrow u \, l \, \nu$ decays}
\label{Vub}

The analysis of 
exclusive charmless semileptonic decays, in particular the  $\bar B \rightarrow \pi l \bar \nu_l$ decay,   is currently employed to determine the CKM parameter \Vub, which
 plays a crucial role in the study  of
the unitarity constraints.
Most studied approaches to calculate the form factors are  once again  lattice QCD (LQCD) and light-cone  QCD sum rules (LCSR), whose domains of applicability are somewhat complementary, lying at high and low $q^2$, respectively.
Very recent \Vub estimates, all in agreement among them,  have been reported by the Babar Collaboration, see Table VII of Ref.  \cite{Lees:2012vv}.
In Table \ref{phidectab2} we  compare two of them, 
an  average estimate  determined
from the simultaneous fit to experimental data
and the LQCD theoretical predictions,  and an estimate obtained using a LCSR determination for the form factor \cite{khos}.
%Other estimates can be found in Table VII of \cite{Lees:2012vv}.

%%%%%%%%%%%%%%
\begin{table}[t]
\centering
\vskip 0.1 in
\begin{tabular}{|l|c|} \hline \hline
{\it \bf  Exclusive Decays}  &    $|V_{ub}|\, (10^{-3})$   \\
\hline \hline
LQCD \cite{Lees:2012vv} &  $ 3.25 \pm 0.31 $ \\
LCSR \cite{Lees:2012vv, khos} & $ 3.46 \pm 0.06 \pm 0.08^{+0.37}_
{-0.32}  $\\
\hline \hline
{\it \bf Inclusive Decays}  &    \\
\hline
\hline
BLNP  & $ 4.40 \pm 0.15^{+0.19}_{-0.21}  $\\
\hline
DGE   & $4.45 \pm 0.15^{+ 0.15}_{- 0.16}$\\
\hline
ADFR   & $4.03 \pm 0.13^{+ 0.18}_{- 0.12}$\\
\hline
GGOU   & $4.39 \pm  0.15^{ + 0.12}_ { -0.20} $\\
\hline
\hline
{\it \bf Global CKM fits}  &    \\
\hline
\hline
CKMfitter  \cite{Derkach}   & $ 3.42^{+0.2}_{- 0.1} $\\
\hline
UTfit \cite{tar}  & $ 3.62 \pm 0.14 $\\
\hline
\end{tabular}
\caption{ Comparison of exclusive, inclusive and indirect  determinations of $|V_{ub}|$. For the exclusive LCSR determination, the  three uncertainties  are statistical, systematic and theoretical, respectively. Inclusive values are taken by HFAG  \cite{Amhis:2012bh}. }
\label{phidectab2}
\end{table}
%%%\end{tabular}
%%%\end{ruledtabular}
%%%\end{table}
%%%%%%%%%%%%%%%%%%%%%%%%%%%%%%%%%%%%%%

The extraction of  $|V_{ub}|$  from inclusive $ \bar  B \rightarrow X_u  l \bar \nu_l$  decays  would follow  in the footsteps of the $|V_{cb}|$ determination, if not for the copious background from the
$ \bar B \rightarrow X_c l \bar \nu_l$ decay.
To overcome this background, inclusive $ \bar  B \rightarrow X_u  l \bar \nu_l$  measurements utilize restricted regions of phase space,
where  the $ \bar B \rightarrow X_c  l \bar \nu_l$    process is  highly suppressed by kinematics. These regions overlap with the threshold one, 
complicating  the
theoretical issues considerably.

It is a long standing problem the discrepancy between the values of $|V_{ub}|$ extracted from inclusive and exclusive decays.
On the experimental side, a lot of effort has  been devoted to enlarge
the  experimental range, so to reduce
on the whole  the weight of the endpoint region.
 Latest results by Belle \cite{Urquijo:2009tp}
 access about the $ 90$\% of the $ \bar B \rightarrow X_u  l \bar \nu_l$ phase space, claiming an overall uncertainty of 7\% on $|V_{ub}|$.
A similar portion of the phase space is covered also by the more recent Babar analysis \cite{Lees:2011fv}.
On the theoretical side several approach have been devised 
to analyze data in the threshold region,  with differences
in treatment of perturbative corrections and the
parameterization of nonperturbative effects.

The latest experimental determinations of $|V_{ub}|$ come from  Babar \cite{Lees:2011fv} and HFAG \cite{Amhis:2012bh} Collaborations.
Both Collaborations extract $|V_{ub}|$  from the partial branching fractions
relying on  at least four different QCD calculations of the partial
decay rate, that is 
 BLNP
by Bosch, Lange, Neubert, and Paz \cite{BLNP1, BLNP2, BLNP3}; DGE, the
dressed gluon exponentiation, by Andersen and Gardi \cite{DGE}; ADFR by Aglietti, Di Lodovico, Ferrara, and Ricciardi
\cite{res4, Aglietti:2006yb,  Aglietti:2007ik}; and GGOU by Gambino, Giordano, Ossola
and Uraltsev \cite{gambino07}.
In Table \ref{phidectab2} we reports the  estimates  by the HFAG \cite{Amhis:2012bh} Collaboration, where  the same inputs have been used for all frameworks; the results are roughly
consistent among them.
Other approaches have been discussed in \cite{Bauer:2001rc, Leibovich:1999xf,Lange:2005qn, SIMBA}.
Notwithstanding all the experimental and theoretical efforts, 
the values of $|V_{ub}|$ extracted from inclusive decays maintain about two $\sigma$ above the values given by exclusive determinations.
Also indirect fits prefer a lower value of $|V_{ub}|$. Recent results from CKMfitter and UTfit Collaborations are  listed in 
Table \ref{phidectab2} as well.

%Long standing |Vub| discrepancy between exclusive, inclusive and UT fits determinations; tensions with SM in some observables (e.g. isospin asymmetry in B → K μ+ μ- at %LHCb…)

%CKM determinations in charm sectors consequential for other processes as well

%Encouraging and impressive recent experimental progresses (notably LHCb and BES III joining the arena)

\section{$ b \rightarrow s (d) \, l^+ \, l^-$ decays}

The  increased luminosity of the actual experimental facilities has made possible 
 to explore flavour-changing
neutral current (FCNC) decays in quantitative detail.
In the SM, FCNC decays  are forbidden  at tree level and  driven by loop diagrams, therefore they
are particularly sensitive  to non-standard virtual contributions. 

In the inclusive $B \rightarrow X_s l^+ l^-$ decays 
the major theoretical uncertainties arise from
the non-perturbative nature of the intermediate  $\bar c c $ states.
By cutting on the invariant dilepton mass around the masses of the $J/\psi$  and  $\psi^\prime$  resonances, 
rather precise determinations seem to be possible, since 
below or above the   $\bar c c $  resonances, 
the inclusive decay is dominated by perturbative contributions.

The calculations of the perturbative contribution has been completed up to next-to-leading (NLO) 
order  in QCD
\cite{QCD1, QCD2, QCD3, QCD4, QCD5}.
In the last 10 years, it has been  extended to the next-to-next-to leading order  (see Ref. \cite{Buras:2011we} and Refs. within)  greatly reducing the theoretical uncertainty, in particular
% the $O(\alpha_s$) corrections to the Wilson coefficient of the operator $(\bar s b)_{V_A} (\bar \mu \mu)_V$  have been completed,  removing 
the large matching scale uncertainty of 16\% at the NLO  level. 
The inclusive $B \ra X_s l^+ l^-$  decays have been  measured at Belle and at Babar \cite{Iwas, Chiang:2010zz, aub1}, and their branching ratios found of  order $10^{-6}$, consistent with SM expectations.
Also a recent  model-independent fit of some  short-distance couplings
shows consistence with the SM \cite{Beaujean:2012uj}.

In the case of inclusive  $B \ra X_d \,  l^+ l^-$, the short distance  analysis is very similar, 
once one keeps the CKM suppressed terms in the operator expansion.
The first $  b \ra d l^+ l^-$ transition has been recently observed in the channel
 $ B^+ \ra \pi^+ \mu^+ \mu^-$  by the LHCb Collaboration
\cite{LHCb2}. The predicted SM branching ratio is of order $10^{-8}$.

The NLO and NNLO QCD corrections for inclusive decays can of course be also
used for the corresponding exclusive decays, that are easier to measure. 
The kinematic available phase space in $ B \ra K^{(\ast)} \mu^+ \mu^-$
 is fully covered experimentally, with the
exception of the $ J/\psi$ 	 and $\psi^\prime$ resonances, which are removed
by cuts.
 A lot of efforts has been devoted to  the determination of the form factors.
In the  exclusive channel $ B \ra K^{\ast} l^+ l^-$, 
it has been shown that a systematic theoretical description using QCD factorization in the heavy quark limit
 is relevant for small invariant dilepton masses and reduces the number of
independent form factors from 7 to 2 \cite{Beneke:2001at}. Spectator effects, neglected in naive factorization, also
become calculable.
The region of low $q^2$,  the dimuon invariant mass squared, where the energy of the emitted meson is large in the
B meson rest frame,  is also the region of applicabity of LCSR
to calculate form factors (see e. g. \cite{ball}).
Also Soft Collinear Effective Theory
(SCET) 
 has been applied at large recoil of the $K^{(\ast)}$ system, typically in the
range between 1 and 6 GeV (for a  recent Ref., see  \cite{Bell:2010mg}).

In 
the high $q^2$ region,
 QCD factorization is
less justified, becoming invalid close to the endpoint of the spectrum at $q^2 = (m_B-m_K)^2$.
Alternative approaches have been developed, based on expansions whose scale is set by the large value of $q^2$ \cite{buch, grin, Beylich:2011aq, Bobeth:2010wg}.
The 
 large $q^2$ region is the domain of election for 
lattice QCD; 
unquenched calculations of form factors  have been recently performed \cite{Liu:2011raa, Zhou:2011be}.
In the same large $q^2$  region, ratios of  $ B \ra K^{\ast}$ form factors have been extracted from angular variables recently measured \cite{ang1, ang2, Albrecht:2012is, Serra:2012mb},  precisely the fraction
of longitudinally polarized vector mesons  and
the transverse asymmetry in the $ B \ra K^{\ast} l^+ l^-$ decay,  and found consistent  with  lattice results \cite{Hambrock:2012dg}.
In general, 
the study of angular observables can be used advantageously in the $ B \ra K^{\ast} l^+ l^-$ decay, even to explore the possibilty of new physics \cite{Becirevic:2012fy, Altmannshofer:2012az, Kosnik:2012dj, DescotesGenon:2012zf}.
The angular distribution
$ B \ra K^{\ast} (\ra K \pi)  l^+ l^-$ may be polluted by events coming from the distribution
$ B \ra K_0^{\ast} (\ra K \pi)  l^+ l^-$, where $K_0^\ast$ is a scalar meson resonance, and this possibility was  analyzed in \cite{Becirevic:2012dp, Matias:2012qz}.

 LHCb  has recently reported the most precise measurement of the branching ratio for the 
$ B^+ \ra K^+  \mu^+ \mu^-$ channel  to date, together with a study of its  angular distribution and differential branching fraction \cite{LHC1}.
In the SM, the differential decay rate  can be written as
\beq
\frac{1}{\Gamma} \frac{d \Gamma \left(  B^+ \ra K^+  \mu^+ \mu^-  \right) } {d \cos \theta} =
\frac{3}{4} \left( 1 -{\mathrm F_H} \right) ( 1- \cos^2 \theta) +  \frac{1}{2} F_H + {\mathrm A_{FB}} \cos \theta
\eeq
 wHere  $\theta$ is  the angle between the $\mu^-$ and the $K^+$ in the rest frame of the
dimuon system. The two parameters, ${\mathrm F_H} $ and  the forward-backward asymmetry of the dimuon system, ${\mathrm A_{FB}} $, depend on  $q^2$. In the SM, ${\mathrm A_{FB}}$ is zero and $F_H$ highly suppressed, and their  measured values are consistent with the SM expectations.
The differential branching fraction of the
$ B^+ \ra K^+  \mu^+ \mu^-$ decay is, however, consistently below the SM prediction at low $q^2$ \cite{LHC1}.
LHCb reports also the actual more precise determinations of  ${\mathrm A_{FB}} $ for the decay $B^0 \ra K^{\ast 0} \mu^+ \mu^-$   \cite{Aaij:2011aa}.

%The decay $B_s  \ra \phi \mu^+ \mu^-$ has been recently observed at CDF \cite{Aaltonen:2011cn}, 

\section{Charm decays}

In the past decade, charm semileptonic  decays have not received the same first-rate attention than beauty decays, but they are rapidly gaining ground.
The extraction of  $|V_{cd}|$ and $|V_{cs}|$ follows in the footsteps of the just described extraction of  $|V_{ub}|$  and $|V_{cb}|$ from semileptonic $B$ decays. Once again, the main theoretical hardship come from the nonperturbative evaluation of the form factors.
Lately, 
 high statistics studies on the lattice have become available. The HPQCD Collaboration  has estimated
the value   $ |V_{cd}| = 0.225 \pm 0.006_{\mathrm{exp}} \pm  0.010_{\mathrm{lat}}$ \cite{Na:2011mc}, with the first error coming from experiments and the second from their lattice computation. The result  is in agreement with the value of $|V_{cd}|$ the same collaboration  has recently
 extracted from leptonic decays and from determinations of $|V_{cd}|$ coming from neutrino scattering.
Instead,
their   best, preliminary value 
   $|V_{cs}| = 0.965 \pm 0.014$  \cite{Koponen:2012fu} shows a discrepancy with the  average value  from leptonic decays
$|V_{cs}| = 1.010 \pm 0.017$
\cite{Rong:2012pb}.
Both $|V_{cs}|$ and $|V_{cd}|$ semileptonic estimates are in agreement with indirect fits.
Other $D$ and $D_s$ decays, such as $D_s \rightarrow \phi \, l \, \nu_l$ or $D \rightarrow \pi \, l \, \nu_l$, have been analyzed as well \cite{Koponen:2012fu}.

Results from QCD light-cone sum rules on $|V_{cs}|$ and $|V_{cd}|$ give  substantial  agreement on the averages and 
 higher theoretical errors with respect to the previous quoted  lattice  results \cite{Khodjamirian:2009ys}.
Recently, a revised version of QCD sum rules reports  reduced errors and  
 an  higher average value
$|V_{cd}|= 0.244 \pm 0.005 \pm 0.003 \pm 0.008 $,  the first and second errors being of an experimental origin and the third  due to the
theoretical uncertainty \cite{Li:2012gr}.

According to  lattice determinations in  \cite{Koponen:2012fu},  the form factors
are insensitive to the spectator quark: $D_s  \rightarrow \eta_s l \nu_l $ and $D \rightarrow K l \nu_l $ form
factors are essentially the same, and the same holds for   $D_s  \rightarrow K l \nu_l $ and $D \rightarrow \pi l \nu_l $ 
 within 5\%.
This result, which  can be tested experimentally,  is expected to hold
also for $B$ meson decays so that $B_s \rightarrow D_s$ and $B \rightarrow  D$ form factors would be equal.

The decays driven by $ c \rightarrow u l^+ l^-$ are forbidden at tree level in the standard model (SM) and proceed by one loop diagram  at leading order in the eletroweak interactions.
Virtual quarks in the loops are of the down type,  and no 
breaking due to the large top mass occurs. The  GIM  mechanism works more effectively  in suppressing FCNC decays than  their strangeness and beauty analogues, leading to tiny decay rates,  dominated by
long distances  contributions. They set the scale, with  branching fractions of order $10^{-6}$, shielding possible enhancements due to new physics.
A way out is to  choose appropriate observables containing mainly short distance
contributions, whose order of magnitude lays way behind $10^{-6}$.

 This year,
effects of possible new physics  have been  been investigated in
$ D^+ \ra \pi^+ \mu^+ \mu^-$ and $D_s^+ \ra K^+ \mu^+ \mu^-$ decays, around the 
$\phi$ resonant peak in spectrum of dilepton invariant mass, concluding that in favourable conditions their value can be as high as 10\% \cite{Fajfer:2012nr}.
Older studies report  investigations of semileptonic decays in the framework of other new physics models,
 such as
 R-parity violating
supersymmetric models,
extra heavy up vector-like quark models  \cite{Fajfer:2007dy},
Little Higgs \cite{Paul:2011ar},
 and leptoquark models \cite{Fajfer:2008tm}. Theoretical analysis of 
semileptonic four 
body decays $ D^0 \ra h_1^+ h_2^- l^+ l^-$, with $l=e, \mu$ and $h_i= \pi, K$, have  been reported this year as well \cite{Bigi:2011em, Cappiello:2012vg}.

\bigskip
I am grateful to the organizers of FPCP12 for inviting me to this very  interesting Conference and for providing  a warm and stimulating atmosphere.

\def\Discussion{
\setlength{\parskip}{0.3cm}\setlength{\parindent}{0.0cm}
     \bigskip\bigskip      {\Large {\bf Discussion}} \bigskip}
\def\speaker#1{{\bf #1:}\ }
\def\endDiscussion{}

\end{document}